\documentclass{aa}
\usepackage{graphicx,psfig}
\usepackage{natbib}
\usepackage{txfonts}
\def\bp{\object{$\beta$\,Pictoris}}

\def\be{\begin{equation}}
\def\ee{\end{equation}}
\begin{document}
\title{Vertical structure of debris discs}

\author{P. Th\'ebault\inst{1,2}}
\institute{
LESIA, Observatoire de Paris,
F-92195 Meudon Principal Cedex, France
\and
Stockholm Observatory, Albanova Universitetcentrum, SE-10691 Stockholm,
Sweden}

\offprints{P. Th\'ebault} \mail{philippe.thebault@obspm.fr}
\date{Received; accepted} \titlerunning{Vertical structure of debris discs}
\authorrunning{Th\'ebault}

\abstract
%
{The vertical thickness of debris discs is often used as a measure of these systems' dynamical excitation, and as clues to the presence of hidden massive perturbers such as planetary embryos. However, this argument might be flawed because the observed dust should be naturally placed on inclined orbits by the combined effect of radiation pressure and mutual collisions.
}
%
{
We critically reinvestigate this issue and numerically estimate what the "natural" vertical thickness of a collisionally evolving disc is, in the absence of any additional perturbing body.
}
{
We use a deterministic collisional code, following the dynamical evolution of a population of indestructible test grains suffering mutual inelastic impacts. Grain differential sizes as well as the effect of radiation pressure are taken into account.
}
{
We find that, under the coupled effect of radiation pressure and collisions, grains naturally acquire inclinations of a few degrees. The disc is stratified with respect to grain sizes, with the smallest grains having the largest vertical dispersion and the bigger ones clustered closer to the midplane.
}
{
Debris discs should have a minimum "natural" observed aspect ratio $h_{min}\sim 0.04\pm0.02$ at visible to mid-IR wavelengths where the flux is dominated by the smallest bound grains. These values are comparable to the estimated thicknesses of many vertically resolved debris discs, as is illustrated with the specific example of AU Mic. For all systems with $h\sim h_{min}$, the presence (or absence) of embedded perturbing bodies cannot be inferred from the vertical dispersion of the disc.
}
\maketitle

\section{Introduction}

\subsection{vertical structure of debris discs}

Dusty debris discs have been observed around main sequence stars for over two decades. It is believed that the detected dust is the lower end of a collisional cascade starting at larger, possibly planetesimal-sized bodies which are steadily eroding due to high velocity impacts \citep{wyatt08}.
As of April 2009, 18 such debris discs have been spatially resolved (see {http://www.circumstellardisks.org/}), mostly in scattered light (visible and near IR) but also in some cases at mid-IR and up to millimetre wavelengths. These images have revealed a great variety of radial and azimutal features, such as rings, clumps or two-side asymmetries, and almost no system displays a featureless structure. These features have been extensively studied, mostly through numerical modelling, considerably improving our understanding of the processes at play in debris discs. Depending on the specificities of each system, they have been interpreted as being the signatures of perturbing planets or stellar companions, or the results of transitory catastrophic events, or due to interactions between the dust and remnant gas.

In comparison, fewer discs have been resolved in the $vertical$ direction. Edge-on seen systems are in this respect the most favourable cases. The best resolved disc so far is $\beta$-Pictoris, for which the vertical structure is fairly well constrained, with a relatively constant thickness $H \sim 15-20\,$AU
(defined by half the vertical FWHM) in the $r<120$\,AU region and an aspect ratio \footnote{the aspect ratio should not be confused with the opening angle $2H/r$} $h=H/r \sim 0.05-0.1$ in the outer regions \citep{heap00,goli06}. Observations in scattered-light have also revealed a striking warped profile and possibly the presence of two slightly tilted distinct discs \citep[e.g][]{moui97,heap00,goli06}. The other relatively well resolved disc is around the $\beta$ Pic coevol star Au Mic. It resembles to some extent a scaled-down $\beta$ Pic, with a roughly flat disc in the inner $r<30\,$AU region and a $H/r \sim 0.05$ beyond that \citep{krist05}, but no obvious vertical asymmetries or warps.
For the two other resolved edge-on systems, HD15115 and HD139664, the vertical structure is less constrained because of poorer resolution and signal to noise ratios. HD15115 has a measured aspect ratio of $\sim 0.05$ but seems to display pronounced asymmetries in its half-width at quarter maximum profile \citep{kalas07}, while HD139664's vertical profile has not been constrained yet \citep{kalas06}. For discs seen head-on or at high angle, the geometrical configuration is less favourable, and the vertical structure is difficult to retrieve, or only through model dependent inversion of brightness profiles. The HD181327 rings for example, might have $H/r$ ratio as large as about 0.1 at the positions of maximum surface density, but the actual ratios could be two times smaller \citep{schnei06}. There is however one non-edge-on system for which a precise thickness estimate has been provided, Fomalhaut, which might have an aspect ratio as low as 0.0125 \citep{kalas05}. We discuss this specific case in more detail in Section 4.3.

Note that there is often an ambiguity regarding the way the thickness $H$ is defined. For an edge-on seen system, this thickness, and all the directly related parameters such as aspect ratio or opening angle, is measured on the disc's luminosity profile along its two radial extensions. In other words, it is $not$ a measure of the disc's local thickness at a given radial location, but an observed thickness, integrating along the line of sight grains from different radial distances \footnote{all those with $r'=r.sin(\theta)$, where $r$ is the distance measured along the ansae and $r'$ the real radial distance}. For systems seen at higher angle, however, the thickness inferred through model dependent fitting is usually the real local thickness at a given radial distance from the star.

\subsection{disc thickness as a measure of dynamical excitation}

Despite the relative lack of observational data, disc vertical structures, or at least their global aspect ratios, have often been used as crucial parameters in debris disc studies, in particular those aimed at modelling their dynamical and collisional evolution. Indeed, disc thickness is commonly considered as being the only observable that can give a direct information about the system's dynamical excitation, and hence about the processes and bodies shaping it. The argument is the following: a system of mutually colliding bodies orbiting around a central object tends toward an equilibrium where impact velocities $\Delta v$ are on average of the order of the bodies surface escape velocity $v_{esc}$. In addition, there is an equipartition between in-plane and vertical motions such as $\langle i \rangle \sim 1/2 \langle e \rangle$, where $i$ and $e$ are the orbital inclinations and eccentricities respectively. This means that if the system is only made of small dust grains, then it should be almost razor thin, with $\langle i\rangle \sim 0$. If on the contrary it has a finite observable thickness, then it means that "something" is dynamically stirring up the system to these large values. This argument is often used to infer the presence of unseen large bodies which are governing the disc's dynamics, and also to quantitatively estimate their size $s_{big}$ from the value of the disc's aspect ratio $H/r$ \citep[e.g.][]{arty97,theb07,quillen07}. A detailed description of this procedure, taking into account several parameters such as the disc's age, surface density, etc., can be found in \citet{quillen07}. We only present here a simplified version, based on the first order assumption that the encounter velocity dispersion $\langle \Delta v \rangle$ is of the order of $v_{esc}(s_{big})$ \citep{arty97}. $s_{big}$ can then be estimated through the equipartition condition $\langle i \rangle \sim 1/2 \langle e \rangle$ coupled to the relation \citep{wethste93,wyatt02}:
\begin{equation}
<\Delta v> \sim \left( <i^2> + 1.25<e^2> \right)^{0.5}{v_{Kep}(r)}
\sim v_{esc}(s_{big})
\label{Rdyn}
\end{equation}
where $v_{Kep}(r)$ is the orbital velocity at radial distance $r$. For a typical debris disc aspect ratio $H/r=0.05$ at a typical distance $r=50$AU from the star, this gives $v_{esc}(s_{big})=400$m.s$^{-1}$, and thus $s_{big} \sim 200-500$\,km, depending on the bodies physical density \footnote{This is a first order estimate. In reality the constraint is not directly on $s_{big}$ but on the product $s_{big}\Sigma_{big}$, where $\Sigma_{big}$ is the surface density of $s_{big}$ bodies \citep[see][for more details]{quillen07}}.

This argument gains additional credibility from the fact that the inferred $s_{big}$ values make sense within the frame of our understanding of debris disc evolution. Indeed, it is believed that the debris disc phase starts when the bulk of the planet formation process is over and large, Ceres to Moon-sized planetary embryos have formed \citep[e.g.][]{ken04}. Observations thus seem to corroborate theory.

However, there could be a major problem with this observation-based derivation of $s_{big}$. Indeed, it overlooks one crucial fact, i.e. that the smallest observable dust grains, those which dominate the flux in scattered light, are strongly affected by radiation pressure from the central star. These particles' eccentricities are thus not solely imposed by the gravitational perturbations of hypothetical large embedded bodies but also by the  eccentricity $e_{RP}$ they naturally gain when being produced. For the simplified case of a body of size $s$ produced from a parent body on a circular orbit, this eccentricity reads
\begin{equation}
e_{RP} = \frac{\beta(s)}{1-\beta(s)}
\label{eRP}
\end{equation}
where $\beta(s) \sim 0.5\times s_{cut}/s$ is the ratio between the radiation pressure force and gravity ($\beta>0.5$ for unbound orbits with $s<s_{cut}$). Once placed on such high-$e$ orbits, these particles will impact other bodies at high velocities. These collisions will necessarily transfer a fraction of the high horizontal (in-plane) velocities into vertical ones. Exactly how much $\Delta v$ is transferred depends on the geometry of the impact, the mass ratio between impactor and target and the distribution of fragments produced by the collision, but it is certain that $some$ in-plane to out-of-plane velocity transfer will occur. As a consequence, the disc should have a natural tendency to thicken, even without any external source of perturbations. If proven significant, this effect could greatly affect the interpretations of debris disc images, by questioning the direct link between disc thickness and dynamical stirring.

In this study, we propose to numerically investigate this issue by quantifying the "natural" vertical structure of a collisionally evolving debris disc, and in particular its equilibrium vertical thickness.

\section{Model} \label{model}

Numerical studies of debris discs fall into two main categories: those investigating the collisional
and size distribution evolution of the system are usually statistical particle-in-the-box models, with no or poor spatial resolution and dynamical evolution \citep[e.g][]{kriv06,theb07,loeh08}, while those studying the dynamics and the formation and evolution of spatial structures are mostly N-body type codes, where size distributions and mutual collisions are usually neglected \citep[see for example ][]{reche08}.

The present problem poses however a specific challenge, inasmuch as it aims to quantify the effect of mutual collisions, within a population of bodies of different sizes, on the dynamical structure. Unfortunately, to study both the dynamical and the collisional (and size) evolution of a dusty disc is far beyond the present day computing capacities, and only a simplified approach is possible. In this respect, one of the most suitable available codes is probably the one developed 3 decades ago by \citet{bra77} and \citet{brahen77} to study the collisional evolution of planetary rings, later upgraded by \citet{thebra98} for the study of collisions among planetesimals. We chose to use here a revised version of this code, adapted to the present problem.

\subsection{code}

It is an N-body type deterministic code, following the dynamical evolution of a population of massless test particles under the influence of a central body's potential as well as possible additional gravitational perturbers. It has a build-in collision search algorithm, which tracks, at each time step, all possible 2-body mutual encounters within the population. Due to the unavoidable limited number of test particles, typically a few $10^4$, the statistics of impacts is too poor with the particles' real sizes. We thus take the usual "inflated radius" assumption, assigning each particle a numerical size $s_{num}$ large enough to get enough impact statistics. Such an approximation is justified as long as $s_{num}$ is smaller than the minimal distance over which dynamical conditions change in the system \citep[e.g.][]{char01}.

Collisions are then treated as inelastic rebounds. Collision outcomes are estimated by computing the momentum before and after impact $p_{bef}$ and $p_{aft}$ in the two-bodies centre of mass frame. The link between $p_{aft}$ and $p_{bef}$ is parameterized by a normal and transversal rebound coefficients, $\epsilon_N$ and $\epsilon_T$, such as $\epsilon_N=-1$ and $\epsilon_T=1$ for a perfectly elastic collision.

For the present problem, we modified the code in order to accommodate for a size distribution within the test particle population, and more specifically for impacts between objects of different sizes. In this case, the momentum balance is weighted by the mass ratio between the 2 impacting bodies. Another crucial improvement is that we now take into account the effect of radiation pressure on the smallest particles. This force is computed by correcting the gravitational force from the central star by a factor $(1-\beta(s))$, where $\beta(s) = F_{RP}/F_{grav}$. In this respect, it might be more convenient to scale particle sizes by their $\beta$ value, with $\beta(s)=0.5$ for the smallest bound objects (if released from circular orbits).

Of course, this "bouncing balls" model is a crude approximation of the real behaviour of a collisionally evolving debris disc. In a real system high-$\Delta v$ collisions should result in fragmentation rather than inelastic rebounds, producing each time a cloud of small fragments. As mentioned before, such chain reactions of fragment-producing collisions are impossible to numerically model. However, if we make the reasonable assumption that a steady-state has been reached in the system, so that the size distribution no longer globally evolves due to collisions \footnote{for each given size there is an equilibrium between the particles lost by fragmenting collisions and the ones newly produced as fragments from collisions involving bigger bodies}, then the present code gives a first order estimate of how mutual collisions between bodies of different sizes redistribute in and out-of-plane velocities. Indeed, this effect is for each impact proportional to the relative velocity, the mass ratio between impactors, the geometry of the impact and energy dissipated at impact, all parameters which our code handles (the energy dissipations being tuned-in by the values for $\epsilon_N$ and $\epsilon_T$). We discuss these issues in more details in Sect.\ref{limits}.

\subsection{setup} \label{setup}

\begin{table}
\begin{minipage}{\columnwidth}
\caption[]{Nominal case set up}
\renewcommand{\footnoterule}{}
\label{init}
\begin{tabular*}{\columnwidth} {ll}
\hline
Radial extent \footnote{initial location of particles' \emph{periastron}} 
& $10<r<100\,$AU\\
Radial surface density profile & $\Sigma \propto r^{-0.5}$\\
Number of test particles $N$ & $5\times10^4$\\
Size range \footnote{as parameterized by $\beta \propto 1/s$} & $\beta(s_{max})=0.025<\beta(s)<\beta(s_{min})=0.4$\\
Normal rebound coef. $\epsilon_N$ & -0.3\\
Transverse rebound coef. $\epsilon_T$ & 1\\
Size distribution & $dN(s) \propto s^{-3.5}ds$\\
Inflated radius (in AU)&  $s_{num}=10^{-4} \times (s/s_{min})$\\
Initial dynamical excitation \footnote{the value for $\langle e_{dyn} \rangle$ is that
of hypothetical large particles with $\beta=0$, not taking into account the eccentricity component due to radiation pressure} & $\langle e_{dyn} \rangle = \langle i\rangle = 0$\\
\hline
\end{tabular*}
\end{minipage}
\end{table}

For the sake of clarity, we consider a nominal reference case, around which several free parameters will be explored individually. This reference case roughly corresponds to a hypothetic typical debris disc, spatially extended from 10 to 100\,AU. We consider $N=5\times 10^{4}$ test particles. The absolute values of their physical radius are not relevant here. Indeed, for the collision outcome procedure, the decisive factor is the impactors mass (or $s^3$) \emph{ratio}. For the collision search routine, the assumed inflated radius is proportional to the physical size $s$, but the proportionality factor, and thus the "real" value of $s$, is of no importance, the only constraint being on the absolute value of $s_{num}$ which has to be large enough impact statistics and small enough not to introduce any bias (see below). As for the orbital evolution computation, what matters is the value of $\beta(s)$ for estimating the radiation pressure force. We thus chose to scale all particle sizes by their $\beta(s)\propto 1/s$ value. We assume that the size distribution has reached a steady state profile following a power law $dN \propto s^{q}ds$, or more exactly $dN \propto \beta^{-q-2}d\beta$. We assume here the classical value $q=-3.5$ for idealized infinite self similar collisional cascades \citep{dohn69}, but another possible values are explored. With such steep size distributions, the size range which can be modelled with $5\times 10^{4}$ particles is necessarily limited. We take here $s_{min}$ and $s_{max}$ such as $\beta(s_{min})=0.4$ (close to the blow-out cut-off size) and $\beta(s_{max})=0.025$. This size range covers however most of the crucial grain population which dominates the flux in scattered light. For the collision search routine, the inflated radius $s_{num}$ is taken proportional to the particles real sizes and is set such as to give enough collision statistics without introducing a bias in impact rate and velocity estimates. We take $s_{num}=10^{-4}\,$AU$\times (\beta(s_{min})/\beta(s)$, which is a typical value for similar impact search routines \citep[e.g.][]{char01,xie08}.

For collision outcomes, we follow \citet{thebra98}, \citet{char01} and \citet{theb01} and take $\epsilon_N=-0.3$ and $\epsilon_T=1$. This corresponds to a $90\%$ energy dissipation for head-on impacts, which is the usual first-order estimate for high velocity impacts \citep[e.g.][]{petfar93,fuji89}, but other values are explored.

In order to avoid confusion, we label $\langle e_{dyn} \rangle$ the "dynamical" component of the particles initial eccentricities. $\langle e_{dyn} \rangle$ is thus a {\it size independent} quantifty, the other component being the (size dependent) radiation pressure imposed eccentricity $e_{RP}$ (see Eq.\ref{eRP}). For the starting dynamical conditions, we consider an extreme fiducial case with $\langle e_{dyn} \rangle = \langle i\rangle= 0$, i.e. the most dynamically quiet state possible for the system. This is in order to get a minimum value for the "natural" thickening effect studied here, i.e., the collisonal transfer of high $\Delta v$ induced by radiation pressure. Note that even if $e_{dyn}=0$, small grains will have a non-zero initial eccentricity because of radiation pressure.

All initial conditions are summarized in Tab.\ref{init}.

We let the simulations run until we do not see any further evolution of the average dynamical characteristics for the different particle size groups. In practice this is achieved once the average number of impacts per particle is $\sim 3$. Beyond this point, the system has reached a steady state and only the slow and progressive decrease of $\langle e \rangle$ and $\langle i \rangle$, due to the energy dissipation at impacts, is observed (see Fig.\ref{inom} and Sec.\ref{limits} for a mode detailed discussion). In this respect, the absolute time scale is not relevant as long as the relaxation time of a few impacts per bodies is short compared to the age of the system, a condition which is most likely to be fulfilled for any observed debris disc \citep[see for instance][]{wyatt05}

\section{Results} \label{res}

\begin{figure}
\includegraphics[angle=0,origin=br,width=\columnwidth]{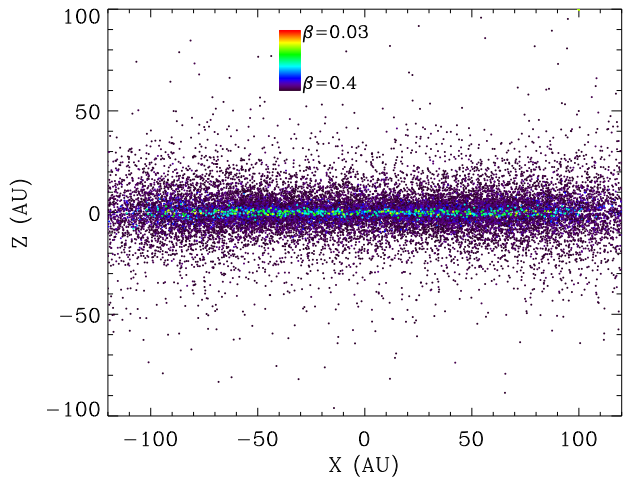}
\caption[]{Final positions in the vertical plane, once the system has reached a collisional steady state, for the $5\times 10^{4}$ particles in the nominal run. The colour scale is indicative of the particles physical sizes.}
\label{nomin}
\end{figure}

\begin{figure}
\includegraphics[angle=0,origin=br,width=\columnwidth]{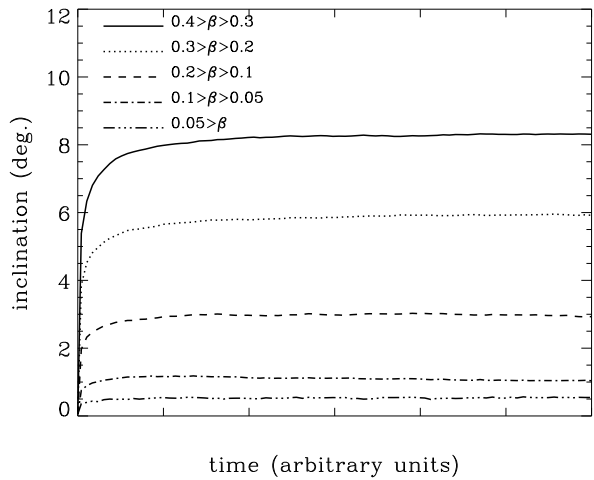}
\caption[]{Evolution of $\langle i \rangle$, for different grain sizes in the nominal run.}
\label{inom}
\end{figure}

Fig.\ref{nomin} shows the final positions, once the system is collisionally relaxed, of all particles in the (x,z) plane. An obvious result is that, in sharp contrast to its initially perfectly thin structure, the disc is now significantly extended in the vertical direction, with a large fraction of the test particles being more than 10\,AU above or below the midplane. Moreover, the disc's vertical structure is strongly stratified with respect to sizes. The bigger objects are clustered close to the midplane while the (most numerous) smaller grains with high $\beta$ have a much larger spread in $z$. This feature appears more clearly in Fig.\ref{inom} showing the evolution of $\langle i \rangle$ for different size ranges: particles close to the blow-out size reach inclinations of about 8$\degr$, but the largest simulated grains (about 15 times this minimum size) settle to less than 0.5$\degr$. These results are logical since the smallest grains are those that have the most eccentric orbits because of radiation pressure. As such they have the highest impact velocities and thus more kinetic energy to transfer from the horizontal to the vertical direction. Likewise, since they are the smallest grains, they are the ones which are most likely to get large dynamical "kicks" from impacts with larger impactors (because of the momentum balance in the centre of frame).

\begin{figure}
\includegraphics[angle=0,origin=br,width=\columnwidth]{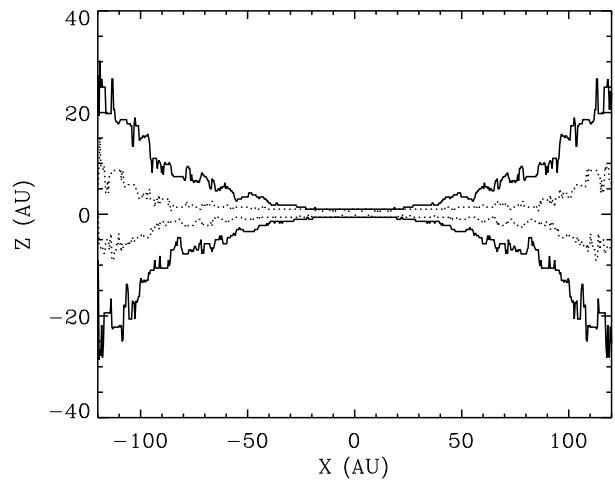}
\caption[]{Synthetic image of a disc seen edge-on, once the steady collisonal state is reached, in scattered light assuming grey scattering. The dotted line corresponds to the vertical full width at half maximum (FWHM), while the full line marks the location of the full width at one tenth maximum FW0.1M}. The fluxes have been averaged over 2AUx1AU bins.
\label{scatter}
\end{figure}
\begin{figure}
\includegraphics[angle=0,origin=br,width=\columnwidth]{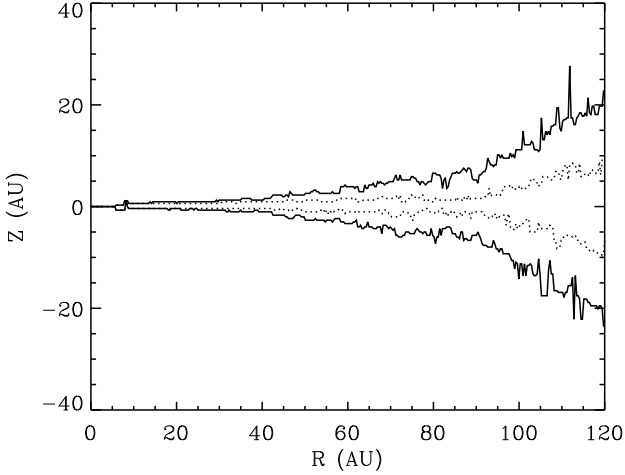}
\caption[]{Same run as in Fig.\ref{scatter}, but displaying the FWHM and FW0.1M for the azimuthally averaged local geometrical thickness as a function of radial distance}
\label{radial}
\end{figure}

To better visualize the effect of this inclination increase on the disc's aspect, we consider two cases: a disc seen edge-on, for which we show a synthetic profile of the disc's vertical FWHM and FW0.1M in scattered light assuming grey scattering (Fig.\ref{scatter}), and a disc seen at higher angle, for which we display the azimuthally averaged $local$ thickness (FWHM and FW0.1M) $H_{geom}(r)$ of the disc, derived from the vertical profile of the integrated geometrical cross section of all particles present in a given radial annulus (Fig.\ref{radial}). Due to the limited number of particles, the resolution is limited to 2AUx1AU and the profiles are relatively noisy, especially in the outermost regions because of the radial and vertical dilution of the particle number density. It is however enough to resolve both the FWHM and the FW0.1M in the whole 0--120\,AU region. As can be seen, the two different thicknesses, i.e., the edge--on profile derived $H_{edge}$ and the local geometrical $H_{geom}$ have relatively comparable values, and can as a first approximation be considered to be identical within the error bars imposed by the noisiness of our profiles. The only differences occur close to the inner edge of the parent bodies disc around 10\,AU, where $H_{geom}$ drops to 0 while the edge-on seen luminosity has a roughly constant thickness in the whole $\sim -20$ to 20\,AU region because it integrates along the line of sight contributions from particles at different radial locations. We ignore for the time being these limited differences (we rediscuss them in Sect.\ref{real}) and consider hereafter that the parameter $H$ stands for both possible definitions of the disc's thickness.

We see that, in both cases, the system's aspect ratio $h=H/r$ is relatively constant and is roughly $\sim$0.04. The FW0.1M profile is much wider, with $h_{0.1} \sim 0.1-0.15$. The seemingly "flatter" aspect of the disc, when compared to the (x,z) positions displayed in Fig.\ref{nomin} is due to the fact that in deriving the scattered light flux, particles' geometrical cross sections are taken into account, which increases the contribution of larger grains, having smaller z dispersion. This is illustrated in Fig.\ref{icomp} showing $\langle i_W \rangle$, the average inclination for the whole grain population weighted by each particle's cross section. It can be seen that $\langle i_W \rangle \sim 3.5\degr$, roughly half that of the smallest grains (Fig.\ref{inom}). Note that this weighted $\langle i_W \rangle$ is approximately equal to $\sqrt(2)h$, which is consistent with the relation between inclination and aspect ratio, for small $\langle i \rangle$ values, in single-sized particle discs \citep{quillen07}.

\subsection{parameter exploration} \label{param}
\begin{figure}
\includegraphics[angle=0,origin=br,width=\columnwidth]{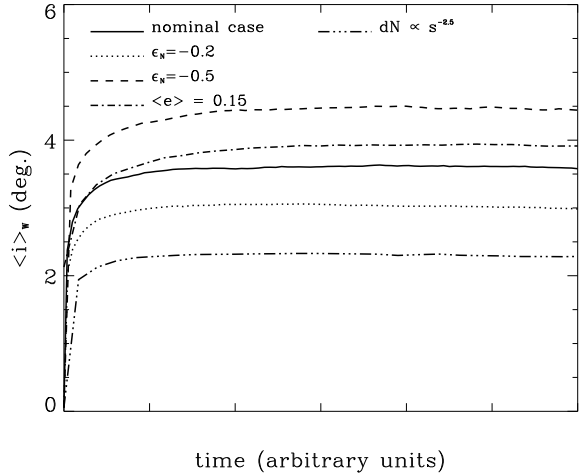}
\caption[]{Evolution of $\langle i_W \rangle$, averaged over all particle sizes, for
different initial set-ups}
\label{icomp}
\end{figure}

Using $\langle i_W\rangle$ as an approximate measure of the disc's vertical
thickness (through the relation $\langle i_W\rangle \sim \sqrt(2)h$), we display
in Fig.\ref{icomp} how our results vary when changing the values of the main parameters considered here.

We first consider two extreme values for normal rebound coefficient $\epsilon_N$, -0.2 and -0.5, corresponding to a 96\% and 75\% energy dissipation respectively, both values being as the upper and lower range of expected energy losses for high velocity fragmenting impacts \citep{fuji89}. As could be logically expected, $\langle i_W\rangle$ decreases with increasing energy dissipation, but the dependence on $\epsilon_N$ remains relatively weak, with $\langle i_W\rangle$ only varying by $\sim 50\%$ between our two extreme cases.

Surprisingly, varying the dynamical excitation of the system doesn't lead to any change in the final $\langle i_W\rangle$ as long as $\langle e_{dyn}\rangle \leq 0.15$. For this range of $\langle e_{dyn}\rangle$, the inclinations acquired by collisional redistribution of the eccentric velocities due to \emph{radiation pressure} (which do \emph{not} depend on $e_{dyn}$) dominate over the value of $\langle i_{dyn}\rangle$.

The only case for which we get a significantly flatter disc is when considering a much shallower slope for the size distribution ($q=-2.5$). In this case, $\langle i_{dyn}\rangle = 2.2^{o}$, almost half the inclination of the nominal case. This is because in this case the system's geometrical cross section, and thus the flux in scattered light, is now dominated by the larger bodies of the distribution which stay on low inclination orbits close to the mid-plane.

\section{Discussion} \label{Discussion}

\subsection{limitations and strength of our approach} \label{limits}

We want to stress that this work is in no way an attempt at realistically modelling the coupled dynamical and collisional evolution of debris discs. As already mentioned, no deterministic code can presently achieve this task, mainly because a correct treatment of collisions would require to take into account the numerous individual small fragments produced after each high-velocity impact, thus leading to an exponential increase of simulated test particles, rapidly exceeding any reasonable value \footnote{Note however the pioneering work of \citet{beaug90}, whose code produced 4 fragments after each shattering collision, but had to be restricted to only 200 initial bodies, or that of \citet{lein05}, allowing the breakup of large rubble piles of gravitationally bound hard spheres, as well as the more recent work by  \citet{grig07} who combined the statistical and N-body approaches by using numerical "super-particles", but were limited to short timescales}. Our study should be regarded as a numerical experiment aimed at exploring one specific mechanism: the ability of mutual collisions to convert the high in-plane velocities due to small, radiation-pressure affected particles into  vertical ones. 

The main and unavoidable problem with our approach is that it considers that particles acquire their dynamics through a succession of inelastic rebounds. This is obviously unrealistic, especially for the smallest grains close to the blow-out size $s_{cut}$. In our runs, such grains evolve mostly through rebounds with grains their own size (which dominate the total population), whereas in reality they should be fragments produced from collisions on bigger parent bodies (while collisions between $\sim s_{cut}$ grains should produce small $<<s_{cut}$ debris quickly removed by radiation pressure). However, for the specific issue addressed here, what matters is only if this departure from reality introduces a systematic error when estimating the final $\langle i \rangle$. Exactly quantifying such a potential bias is clearly beyond the scope of this study, but it is possible to get a first idea. Statistical collisional evolution model \citep[e.g.][]{theb07} show that, for extended debris discs with low dynamical excitation ($\langle e \rangle_{dyn}\leq 0.05)$, grains close to $s_{cut}$ are mostly produced through cratering and shattering collisions between $\sim s_{cut}$ projectiles \footnote{this is mainly because grains close to the blow out size are the impactors carrying the highest kinetic energy and have thus an increased excavating/shattering capacity \citep[see ][ for a detailed discussion on these issues]{theb07}} and targets in the $\sim 5\times s_{cut}$ to $\sim 100\times s_{cut}$ range. In other words, the most common way to produce an $s_{cut}$ fragment is by an impact involving a previous (and ultimately shattered) $\sim s_{cut}$ projectile. If we now use the simplified estimate of \citet{stern97} for the velocity of fragments produced after an impact at speed $\Delta v$ \footnote{This is of course a very simplified expression, which neglects the fact that $v_{fr}$ should in principle depend on fragment size and on the way fragments are produced: cratering or shattering of the target. Nevertheless, we can ignore these refinements for the order-of-magnitude purpose of our discussion}
\begin{equation}
v_{fr} = \left( 2\frac{f_{ke}E_{kin}}{M_{target}}\right)^{0.5}
\label{vfrag}
\end{equation}
where $E_{kin}=0.5(M_{target}+M_{proj})\Delta v^{2}$ and $f_{ke}$ is the fraction of kinetic energy not dissipated at impact, we see that $v_{fr} \sim f_{ke}^{0.5}\Delta v$ when $M_{target}>>M_{proj}$. This means that the fragment velocity is comparable to the outcome velocity for an inelastic rebound between two equal-sized bodies dissipating $(1-f_{ke})$ of the kinetic energy at impact. Thus, by tuning in the value of $f_{ke}$, the dynamical outcome of a collision between two $s_{cut}$ particles bouncing at speed $\Delta v$ can be made to roughly mimic that of a $s_{cut}$ fragment produced by a $\Delta v$ impact of an $s_{cut}$ projectile on a larger target. This is why the rebound coefficients $\epsilon_N$ and $\epsilon_T$ have been given values corresponding to the typical energy loss expected from high velocity fragmenting impacts (see Sec.\ref{setup}). We are nevertheless aware that it is impossible to make this N-body code fully bias free in this respect. It is in particular likely that it slightly overestimates the global $E_{kin}$ transfer for small $\sim s_{cut}$ grains, since a small fraction of them should in reality be produced by fragmenting collisions involving projectiles $>s_{cut}$, for which the impact kinetic energy is lower than that for $\sim s_{cut}$ projectiles.

A second issue is that, due to the succession of dissipative collisions, the numerical system's total kinetic energy is steadily decreasing. In particular, the high initial $\langle e \rangle$ of the smaller particles are progressively damped as they suffer more and more impacts. This is an artificial behaviour, since in reality the smaller grains should not survive several high-$\Delta v$ impacts and should be destroyed before having had the time to be dynamically damped, while newly produced grains of the same size should progressively enter the system with high initial eccentricities ("automatically" acquired when they are collisionally released from their parent body). To minimize this numerical bias, simulations are stopped before the average number of collisions per body exceeds 3. As can be seen from Fig.\ref{inom}, this is enough for the system to reach a steady state in terms of $\langle i \rangle$ for each grain population. Nevertheless, an artificial decrease of the system's excitation is observed, which leads to underestimating the final $\langle i \rangle$.

We thus see that it is impossible to make our N-body approach fully bias free. Nevertheless, for the reasons just discussed, the two main sources of systematic errors should remain relatively limited. Moreover, they should to some extent compensate each other.
As a consequence, we are relatively confident that, for systems which are sufficiently evolved to have reached a steady state where the size distribution and dynamical structure no longer evolves due to collisions, the present code gives a satisfying first-order estimate. As mentioned in Sect.\ref{model}, its main strength is that it takes into account the main parameters that drive the in-to-out-of-plane kinetic energy transfer: collision velocity, impact geometry, mass ratio between
the impactors and energy loss at impact. Note that such a "bouncing balls" approach has also been recently proposed by \citet{chiang09} as the best way to address a not--so--different issue, i.e., the collisional relaxation of grain orbits in the Fomalhaut ring.

\subsection{the "natural" thickness of debris discs} \label{natural}

Despite the limitations inherent to our numerical approach (see previous section), we believe our main result to be relatively robust, i.e., there is a "natural" thickening of a debris disc due to the combined action of radiation pressure and mutual collisions. Expressing the disc's thickness in terms of its average aspect ratio $h=H/R$, we find that (from Fig.\ref{icomp} and using the relation
$\langle i_W\rangle \sim \sqrt(2)h$ )
\begin{equation}
h_{min} = 0.04 \pm 0.02
\label{hrange}
\end{equation}
where the error bar is probably overestimated, as it is obtained when assuming rather extreme values for our parameter exploration (see Sec.\ref{param}). A direct consequence is that there should be in principle no debris disc with an aspect ratio smaller than $h_{min}$, at least in the generic case of systems whose geometrical cross section is dominated by the smallest grains close to the blow-out size (see discussion ot the end of this subsection).

Another important result is that the disc's vertical thickness is almost independent of its intrinsic dynamical excitation as long as $\langle e_{dyn} \rangle=2\,\langle i_{dyn} \rangle \leq 0.15$. In this case, all intrinsic dynamical effects, in particular the out of plane excursion due to the "dynamical" $i$, are masked by the "natural" thickening effect studied here. Likewise, it means that for a disc with an observed $h\leq 0.04$ (or 0.06 to be conservative), the correlation between vertical structure and intrinsic dynamical excitation is lost. In particular, no information on potential embedded perturbing objects can be retrieved, except that they cannot be larger than $\sim 500\,$km (Eq.\ref{Rdyn}). 

A third result is that discs should be stratified with respect to sizes, with the bigger grains being clustered much closer to the midplane. This might have interesting observational consequences. Indeed, it should imply that discs should appear much thinner at longer wavelengths, where smaller grains are poor scatterer/emitter and where the flux is dominated by bigger particles. For a typical value $s_{cut}\sim 2-10\mu$m, this should thus happen for $\lambda \geq 50\mu$m, i.e. in the mid to far IR thermal emission. Unfortunately high-resolution images are very difficult to obtain at these long wavelengths and there is to our knowledge no debris disc that is vertically resolved beyond the mid-IR. 

There might however be one marginal possibility for a disc to appear flatter than $h_{min}$ even in scattered light, i.e. if its total geometrical cross section is dominated by grains significantly larger than the blow out size. This is clearly not the case for a Dohnanyi size distribution in $dN \propto s^{-3.5}ds$, as well as for most of the more realistic "wavy" distributions derived from collisional evolution models \citep[e.g.][]{kriv06,theb07}. There is however a range of systems for which these numerical studies have shown that they could be depleted in small grains, i.e., those with very low dynamical excitation $\langle e_{dyn} \rangle$. Indeed, as identified by \citet{theb08}, for $\langle e_{dyn} \rangle < 0.01$, there is an imbalance between the rates of small grain \emph{production}, which strongly decreases with $e_{dyn}$, and \emph{destruction}, which only moderately changes for low $e_{dyn}$ values (because it is mostly imposed by $e_{PR}$). For these systems, the scattered light flux should be dominated by grains typically 10-100$s_{cut}$ in size, which should have very low $i$ and thus lead to a very flat aspect. Note however that the \citet{theb08} estimates did not take into account the mechanism identified here of collisional redistribution of in-plane to out-of-plane velocities, which should tend to reduce the rate of small grains destructions by spatially diluting them in $z$. It remains to see to which extent this effect might reduce the depletion of $\sim s_{cut}$ grains and if these dynamically cold, small-grains-depleted systems are a realistic alternative. 

Let us stress that this discussion regards the global aspect ratio or thickness of the disc, not possible local clumps, blobs or other asymmetries in its vertical structure, like the famous beta Pic warp. Such local features could well be the signature of hidden perturbers but are not the purpose of the present work.

\subsection{comparison to real systems} \label{real}

\begin{figure}
\includegraphics[angle=0,origin=br,width=\columnwidth]{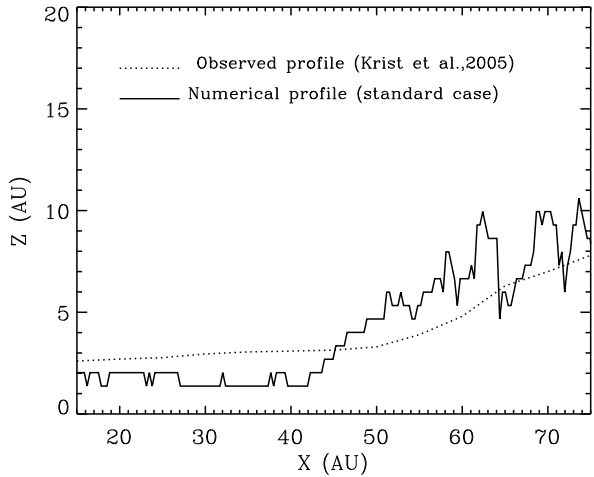}
\caption[] {Vertical FWHM (=$2H$) for the AU Mic disc, as derived from observations \citep{krist05} and obtained for a test numerical disc initially confined to a narrow annulus at $\sim 40\,$AU with  $\langle e_{dyn} \rangle = \langle i_{dyn} \rangle = 0$. The observed FWHM has been averaged over the two ansae \citep[see Fig.7 of][]{krist05}. The synthetic profile has been degraded to the resolution of the scattered light images, i.e. 2\,AU}
\label{aumic}
\end{figure}

As already mentioned, the two best resolved debris discs, \bp\ and AU Mic \footnote{AU Mic is an M star where, in addition to radiation pressure, stellar wind is also acting to place small grains on eccentric or unbound orbits \citep{wood02,aug06,stru06,fitz07}. This additional force has to a first order the same dependency on particle size and radial distance as radiation pressure.}, have aspect ratios $h\sim 0.05$-0.1 and 0.05 respectively. AU Mic's vertical thickness thus falls within the range for the "natural" $h_{min}$ derived in Eq.\ref{hrange}, whereas \bp\ is around the upper end of this range. Note that the preliminary estimate for HD15115 aspect's ratio, i.e. $\sim 0.05$, also falls within our estimated range for $h_{min}$.

To push our analysis a step further we perform one additional run specifically aimed at fitting AU Mic (\bp\ has a much more complex radial and vertical structure, especially its warp, which does not make it an ideal candidate for such a simple fit). For this task, we take input parameters adapted to the specificities of this system, i.e., a "birth ring" (where all non radiation affected larger particles are located) centered at $\sim 40\,$AU and of width $\sim 5\,$AU \citep{aug06,stru06}. As for our previous runs, we consider a limiting dynamically cold case with $\langle e_{dyn} \rangle = \langle i_{initial} \rangle = 0$, the rest of the parameters are those for our nominal set-up (Tab.\ref{init}). As can be seen in Fig.\ref{aumic}, a reasonably good fit of the system's FWHM radial profile is obtained. Note that this simple fit has been obtained with our arbitrarily defined nominal set-up, and we expect much better fits to be possible when tuning in all free parameters (rebound coefficients, size distribution, etc..). Of course, this result should not be regarded as a proof that the AU Mic disc is completely dynamically cold. It is just a numerical test showing that this system's vertical structure is easily explained by the natural thickening mechanism identified in this study, without the need of additional exciting sources such as hidden large planetesimals or planets. The presence of such hidden objects can of course not be ruled out, and they might even make sense within the frame of our current understanding of debris discs, but we want to stress that nothing can be inferred about their presence or absence from the disc's observed aspect ratio.

There is however one specific system which might appear as a counter example to our study: Fomalhaut and its striking debris ring surrounding a recently discovered giant planet \citep{kalas08}. Indeed, this ring's aspect ratio has been observationally estimated to be $h=H/r \sim 0.0125$ \citep{kalas05}, a value which was corroborated by the numerical estimate $h \sim 0.017$ by \citet{chiang09}. These values are below the minimum value for $h_{min}$ derived in Equ.\ref{hrange} and seem to invalidate our conclusions. However, two important points should be stressed here. The first one is that the disc's thickness estimates were obtained for one specific region, i.e., the sharp inner edge of the bright ring located at 133\,AU \footnote{They are based on the measure of this edge's drop-off in surface brightness, which, because of the system's inclination to the line of sight, could either be due to an infinitely flat disc with a finite spread in the radial direction, or a perfectly sharp radial edge with a finite vertical thickness \citep[see Sect.3.2.2 of][]{chiang09}}. The second one is that the thickness which has been estimated for Fomalhaut is the local geometrical vertical thickness $H_{geom}$, and $not$ the line of sight integrated thickness of the luminosity profile as for an edge-on seen disc. We have seen in Sect.\ref{res} that, although these two different thicknesses do most of the time give similar values, $H_{geom}$ distinguishes itself in one specific case, i.e., the inner edge of the parent bodies disc where it can reach very small values (see Fig.\ref{radial}). Exactly how small is difficult to evaluate because of the limited radial resolution of our synthetic profiles, but it is clearly below $H_{min}$, and could thus be compatible with the Fomalhaut estimate. Clearly, this specific case has to be investigated in more details. It would in particular be of great help to obtain a thickness estimate further out in the disc, where our model predicts a wider vertical dispersion.

\section{Summary and Conclusions}

The main results of our numerical investigation can be summarized as follows:
\begin{itemize}
\item Even in the absence of perturbing bodies, debris disc should naturally thicken because of the coupled effect of radiation pressure and mutual grain collisions
\item The basic mechanism is the following: because of radiation pressure, the smallest grains have high in-plane velocities which are partially converted into vertical motions by collisions with other grains. The dynamical outcome depends on the grain differential sizes and the collision geometry.
\item The disc is stratified in the vertical direction with respect to particle sizes, the smallest grains having the largest dispersion while the bigger ones remain close to the midplane.
\item At wavelengths were the flux is dominated by the smallest grains, i.e., in the visible to mid-IR, 
this natural thickening effect places a minimum value $h_{min}$ on a disc's aspect ratio. We find $h_{min}\sim0.04 \pm 0.02$, although this estimate should be taken with some care given the unavoidable limitations of our model.
\item Most vertically resolved debris discs have aspect ratios compatible with $h_{min}$, which can thus be explained without resorting to the perturbing presence of large "hidden" bodies. As an illustration, we obtain a relatively satisfying fit of AU Mic's vertical profile.
\item 
A corollary is that, for all systems with $h \sim h_{min}$, no information about the presence of embedded planetary embryos can be directly retrieved from measuring the disc's vertical dispersion. 
\end{itemize}
Let us conclude by stressing that these results should in no way be interpreted as a proof that debris discs are dynamically cold and devoid of hidden massive perturbing bodies. Indeed, the presence of such perturbers is fully compatible with our results. As an example, a disc with $h=0.04$ could harbour embryos as large as a few 100\,km. As a matter of fact, we subscribe to the theoretical arguments supporting the presence of large hidden bodies, which are in particular needed to provide the mass reservoir for sustaining the high collisional activity of debris discs \citep{loeh08}. What we have shown is that direct observational evidence can probably not be easily obtained, since for aspect ratios lower than $\sim0.06$ there is no direct link between vertical thickness and the dynamical excitation of the system.

{}
\clearpage


\begin{thebibliography}{}
%
\bibitem[Artymowicz(1997)]{arty97} Artymowicz P., 1997, Ann. Rev. Earth Planet. Sci. 25, 175
%
\bibitem[Augereau \& Beust(2006)]{aug06} Augereau, J.-C., Beust, H., 2006, A\&A, 455, 987
%
\bibitem[Beauge \& Aarseth (1990)]{beaug90} Beauge, A., Aarseth, S.J., 1990, MNRAS, 245, 30
%
\bibitem[Brahic(1977)]{bra77} Brahic, A., 1977, A\&A, 54, 895
%
\bibitem[Brahic \& H\'enon(1977)]{brahen77} Brahic, A., H\'enon, 1977, A\&A, 59, 1
%
\bibitem[Charnoz et al.(2001)]{char01} Charnoz, S., Th\'ebault, P., Brahic, A., 2001, A\&A, 373, 683
%
\bibitem[Chiang et al.(2009)]{chiang09} Chiang, E., Kite, E., Kalas, P., Graham, J.R., Clampin, M., 2009, ApJ, 693, 734
%
\bibitem[Dohnanyi(1969)]{dohn69} Dohnanyi J.S., 1969, JGR 74, 2531
%
\bibitem[Fitzgerald et al.(2007))]{fitz07} Fitzgerald, M. P.; Kalas, P. G.; Duchêne, G.; Pinte, C.; Graham, J. R., 2007, ApJ, 670, 536
%
\bibitem[Fujiwara et al.(1989)]{fuji89} Fujiwara, A., Cerroni, P., Davis, D., Ryan, E., di Martino, M., in Asteroids II, eds. R.P. Binzel, T. Gehrels \& M.S. Matthews (Tucson: Univ. of Arizona Press), 1989, 240
%
\bibitem[Golimowski et al.(2006)]{goli06} Golimowski, D. A.; Ardila, D. R.; Krist, J. E.; and 40 coauthors, 2006, AJ, 131, 3109
%
\bibitem[Grigorieva et al.(2007)]{grig07} Grigorieva, A., Artymwicz, P., Th\'ebault, P., 2007, A\&A, 461, 537
%
\bibitem[Heap et al.(2000)]{heap00} Heap, S. R.; Lindler, D. J.; Lanz, T. M.; Cornett, R. H.; Hubeny, I.; Maran, S. P.; Woodgate, B., 2000, ApJ, 539, 435
%
\bibitem[Kalas et al.(2005)]{kalas05} Kalas, P., Graham, J., Clampin, M., 2005, Nature, 435, 1067
%
\bibitem[Kalas et al.(2006)]{kalas06} Kalas, P., Graham, J. ; Clampin, M.; Fitzgerald, M., 2006, ApJ, 637, 57
%
\bibitem[Kalas et al.(2007)]{kalas07} Kalas, P., Fitzgerald, M., Graham, J. , 2007, ApJ, 661, 85
%
\bibitem[Kalas et al.(2008)]{kalas08} Kalas, P.; Graham, J.R.; Chiang, E.; Fitzgerald, M. P.; Clampin, M.; Kite, E. S.; Stapelfeldt, K.; Marois, C.; Krist, J., 2008, Science, 322, 1345
%
\bibitem[Kenyon \& Bromley (2004)]{ken04} Kenyon, Scott J.; Bromley, B., 2004, ApJ, 602, L133
%
\bibitem[Krist et al.(2005)]{krist05} Krist, J.E., et al.\ 2005, AJ, 129, 1008 
%
\bibitem[Krivov et al.(2006)]{kriv06} Krivov, A., L\"ohne, T., Sremcevic, M., 2006, A\&A, 455, 509
%
\bibitem[Leinhardt \& Richardson (2005)]{lein05} Leinhardt, Z., Richardson, D., 2005, ApJ, 625, 427
%
\bibitem[Lissauer \&\ Stewart(1993)]{lisste93} Lissauer J., Stewart G., 1993, in {\it Protostars and Planets III}, the Univ. of Arizona Press, Tucson, 1061
%
\bibitem[L\"ohne et al.(2008)]{loeh08} L\"ohne, T., Krivov, A., Rodmann, J., 2008, ApJ, 673, 1123
%
\bibitem[Mouillet et al.(1997)]{moui97} Mouillet, D.; Larwood, J. D.; Papaloizou, J. C. B.; Lagrange, A. M., 1997, MNRAS, 292, 896
%
\bibitem[Petit \& Farinella(1993)]{petfar93} Petit, Jean-Marc; Farinella, Paolo, 1993, CeMDA, 57, 1
%
\bibitem[Quillen et al.(2007)]{quillen07} Quillen, A.C., Morbidelli, A., Moore, A., 2007, MNRAS, 380, 1642
%
\bibitem[Reche et al.(2008)]{reche08} Reche, R.; Beust, H.; Augereau, J.-C.; Absil, O., 2008, A\&A, 480, 551
%
\bibitem[Schneider et al.(2006)]{schnei06} Schneider, G.; Silverstone, M. D.; Hines, D. C.; Augereau, J.-C.; Pinte, C.; Ménard, F.; Krist, J.; Clampin, M.; Grady, C.; Golimowski, D.; Ardila, D.; Henning, T.; Wolf, S.; Rodmann, J., 2006, ApJ, 650, 414
%
\bibitem[Stern \& Colwell (1997)]{stern97} Stern, A., Colwell, J., 1997, ApJ, 480, 879
%
\bibitem[Strubbe \& Chiang(2006)]{stru06} Strubbe, L.E., Chiang, E.I., 2006, ApJ, 648, 652
%
\bibitem[Th\'ebault \& Augereau (2007)]{theb07} Th\'ebault, P., Augereau, J. C.,  2007, A\&A, 472, 169
%
\bibitem[Th\'ebault \& Beust(2001)] {theb01} Th\'ebault, P., Beust, H. 2001, A\&A, 376, 621
%
\bibitem[Th\'ebault \& Brahic(1998)] {thebra98} Th\'ebault, P., Brahic, A., 1998, Planet. Space Sci., 47, 233
%
\bibitem[Th\'ebault \& Wu(2008)]{theb08} Th\'ebault, P., Wu, Y., 2008, A\&A, 481, 713
%
\bibitem[Wetherill \& Stewart(1993)]{wethste93} Wetherill, G. W.; Stewart, G. R., 1993, Icarus, 106, 190
%
\bibitem[Wood et al.(2002)]{wood02} Wood, B.E.; Müller, H.-R.; Zank, G.P.; Linsky, J. L., 2002, ApJ, 574, 412
%
\bibitem[Wyatt (2005)]{wyatt05} Wyatt, M. C.; 2005, A\&A, 433, 1007
%
\bibitem[Wyatt (2008)]{wyatt08} Wyatt, M. C.; 2008, ARA\&A, 46, 339
%
\bibitem[Wyatt \& Dent(2002)]{wyatt02} Wyatt, M. C.; Dent, W.R.F., 2002, MNRAS, 334, 589
%
\bibitem[Xie \& Zhou(2008)]{xie08} Xie, Ji-Wei; Zhou, Ji-Lin, 2008, ApJ, 686, 570
%
\end{thebibliography}
\end{document}